\documentclass[12pt]{article}
\usepackage[dvips]{graphicx}
\usepackage{latexsym}
\newcommand{\beq}{\begin{equation}}
\newcommand{\eeq}{\end{equation}}
\newcommand{\Frac}[2]{\frac{\displaystyle #1}{\displaystyle #2}}

\def\mapright#1{\smash{
     \mathop{\longrightarrow}\limits^{#1}}}

\begin{document}
\thispagestyle{empty}
\begin{titlepage}
\begin{center}
\vspace*{2.75cm} 
\begin{Large}
{\bf Semileptonic decays of charmed mesons} \\ {\bf in the effective 
action of QCD} \\[2.4cm]
\end{Large}
{ \sc G. Amor\'os$^1$, S. Noguera$^{1,2}$} 
\ and { \sc J. Portol\'es$^1$}\\[0.8cm]

$^1${\it Departament de F\'\i sica Te\`orica, IFIC, Universitat de Val\`encia -
CSIC\\
 Apt. Correus 22085, E-46071 Val\`encia, Spain }\\[0.5cm]
$^2${\it Departament de F\'\i sica Te\`orica, Universitat de Val\`encia, \\
C/ Dr. Moliner, 50, E-46100 Burjassot (Val\`encia), Spain}\\[2.5cm]

\begin{abstract}
\noindent
Within the framework of phenomenological Lagrangians we construct the
effective action of QCD relevant for the study of 
semileptonic decays of charmed mesons. Hence we evaluate the form factors of
$D \rightarrow P(0^-) \ell^+ \nu_{\ell}$ at leading order in 
the $1/N_C$ expansion and, by demanding their QCD--ruled
asymptotic behaviour, we constrain the couplings of the Lagrangian. The
features of the 
model--independent parameterization of form factors provided and their
relevance for the analysis of experimental data are pointed out.
\end{abstract}
\end{center}
\vfill
\hspace*{1cm} PACS~: 13.20.-v, 13.20.Fc, 11.40.-q, 12.38.-t \\
\hspace*{1cm} Keywords~: Semileptonic decays, QCD constraints, form factors
of currents.
\eject
\end{titlepage}

\pagenumbering{arabic}

\section{Introduction}
\hspace*{0.5cm}Matrix elements of hadron currents in exclusive
processes provide, from a 
phenomenological point of view, a detailed knowledge on the hadronization
mechanisms. Their evaluation, however, is a long--standing problem due
to the fact that involves strong interactions in an energy region where
perturbative QCD is unreliable. Within this frame, exclusive semileptonic
decays of mesons
yield the relevant physical system to analyse matrix elements of flavour
changing currents.
\par
When only light quark flavours are involved, as in $K_{\ell 3}$ or 
$K_{\ell 4}$ processes, the model--independent rigorous framework of
Chiral Perturbation Theory ($\chi PT$) allows a thorough study that 
has been proven
successful \cite{kln}. Semileptonic decays of B mesons, on the other side,
can be studied within the Heavy Quark Effective Theory (HQET). This
last procedure relies in the fact that, the $b$ quark being much heavier
than $\Lambda_{QCD}$ (which determines the typical size of hadrons), the 
light degrees of freedom interact independently of the flavour or spin
orientation of the heavy quark. In practice one expands the amplitudes
in inverse powers of the heavy quark mass ($\Lambda_{QCD} / M_b$), and 
the expansion is most suitable for weak decays where heavy flavours
are involved, i.e. $b \rightarrow c$, \cite{hqs}.
\par
However charmed mesons decay to light flavours and the $c$ quark is much
lighter than the $b$ quark; therefore and though the HQET
has also been applied to the study of its semileptonic 
decays \cite{Ca294,Casa106}, involving already a rather 
cumbersome effective action
at the next--to--leading order,
it is doubtful that perturbative corrections are small enough
to provide a thorough result. Another approach involves a mixed
framework including HQET and modelizations \cite{IT195} that, although
predictive, rely in ad hoc assumptions not well justified. 
In addition there is no $\chi PT$ framework appropriate to perform 
this task either because the $c$ quark does not belong to its
realm. This no--man's--land position of charm has
brought about a feeble status in the study of its decays and, in particular,
of $D_{\ell 3}$ semileptonic decays we are interested here. Several
analyses exist within lattice QCD \cite{latq}, QCD sum rules \cite{qsr}, 
and models using phenomenological approaches \cite{italo} or quark
realizations~\cite{isgo}. Sideways non--leptonic decays of charmed mesons
that, up to present, have only been studied in several modelizations such
as factorization \cite{facto} or chiral realizations \cite{yo2}, rely 
within these models in semileptonic form factors. Consequently their study
is also relevant for those processes.
\par
From an experimental point of view, while branching ratios are rather
well measured in both $D \rightarrow P \ell^+ \nu_{\ell}$ and
$D \rightarrow V \ell^+ \nu_{\ell}$ processes \footnote{If unspecified, 
$P$ is short for pseudoscalar meson, $V$ for vector meson, and 
$D$ is short for $D^{+,0}$ or $D_{\textsf{\footnotesize{s}}}^{+}$.
 Charge conjugate modes are
also implied.} \cite{pdg}, the structure
of their form factors, relying more on the statistics of events, is loosely
known \cite{revo}. The E687 and E791 
experiments at Fermilab \cite{oldies,old4,old3}, BEATRICE at CERN \cite{old2},
and CLEO at Cornell \cite{cleito1,cleito2,cleito3} have published their 
analyses and
a further improvement will
continue with FOCUS (E831) in the near future, with approximately forty
times the previous E687 number of events \footnote{Private communication
received from Will Johns.}. Hence form factors in these processes 
are expected to be thoroughly studied.
\par
Effective actions of the underlying Standard Model, as $\chi PT$
or HQET, have become excellent frameworks
to carry on analyses of processes which relevant physics properties are
embodied in phenomenological Lagrangians that contain the proper degrees
of freedom and symmetries. The hadronic system we pretend to describe 
here involves charmed mesons and light pseudoscalar mesons or vector 
resonances. The construction of phenomenological Lagrangians \cite{steve,cole}
gives us a rigorous path to follow when both, Goldstone bosons (light 
pseudoscalar mesons) and matter fields (we include here vector resonances
and charmed pseudoscalar mesons) are involved. In addition we will 
implement this formulation
with suited dynamical assumptions based on large number of colours ($N_C$)
properties \cite{HOO02} and the asymptotic behaviour of QCD. These tools
have largely been employed together with the construction of phenomenological
Lagrangians in order to provide an effective action of the underlying strong
interacting field theory in the non--perturbative, resonance dominated,
energy region.  This procedure has been
successfully applied to the construction of the Resonance Chiral Theory
\cite{toni} providing an excellent basis to parameterize and explore 
the relevant phenomenology.
\par
Within this frame the goal of this paper is to provide a 
model--independent QCD--based parameterization of form factors suitable
for the analyses of the foreseen new data. To go ahead we 
will construct in Section 2 the relevant effective action of QCD for 
the study of
semileptonic decays of charmed mesons. Then we will use this action 
to evaluate the form factors of $D \rightarrow P \ell^+ \nu_{\ell}$
processes in Section 3 and we will impose the 
constraints that the QCD--ruled 
asymptotic behaviour of form factors demand on the coupling constants,
completing therefore the construction of
the effective action. This procedure gives 
a general constrained parameterization of form factors that 
relies on symmetry properties of the underlying QCD theory without
appealing to model--dependent simplifying assumptions.
In the following Section 4 we will comment on the phenomenology and use of our 
results in order to analyse the experimental data of 
$D \rightarrow P \ell^+ \nu_{\ell}$ decays. The complete
study of the $D \rightarrow V \ell^+ \nu_{\ell}$ processes will be
carried on in a later publication \cite{noi}. 
In Section 5 the relevance of semileptonic
processes in determining the couplings of the effective action is pointed
out.
A comparison of our results with those based in the heavy quark
mass expansion will be sketched in Section 6 and, finally,
Section 7 is devoted to underline our conclusions. 

\section{The effective action}
\hspace*{0.5cm} The present construction of effective field theories
of the Standard Model in different energy regions is based in the 
theorem put forward by Weinberg in Ref.~\cite{wein} that, schematically,
says that the most general Lagrangian containing all terms consistent
with the demanded symmetry principles provides general amplitudes 
with the basic properties of a Quantum Field Theory.
\par
Massless QCD with three flavours has a spontaneously broken chiral
symmetry that manifests in the chiral Lagrangian where Goldstone fields
$\varphi_i$ parameterize the element $u(\varphi)$ of the coset space
$G/H \, \equiv \, SU(3)_L \otimes SU(3)_R / SU(3)_V $ given by
\begin{eqnarray} \label{eq:uno}
u(\varphi) \; & \; = \; & \; \exp \left( \, \Frac{i}{\sqrt{2} \,  F}
 \Pi (\varphi) \, 
\right) \; \; , \nonumber \\
& & \\
\Pi (\varphi) \; & \; =  \; & \left( \, \begin{array}{ccc}
                     \Frac{\pi^0}{\sqrt{2}} + \Frac{\eta_8}{\sqrt{6}} &
		     \pi^+ & K^+ \\
		     & & \\
		     \pi^- & - \, \Frac{\pi^0}{\sqrt{2}} + 
		     \Frac{\eta_8}{\sqrt{6}} & K^0 \\
		     & & \\
		     K^- & \overline{K^0} & - \, 2 \, 
		     \Frac{\eta_8}{\sqrt{6}}
		     \end{array} \, 
		     \right)
\; \; , \nonumber
\end{eqnarray}
where $F \simeq 92.4 \, \mbox{MeV}$ is the pion decay constant.
The transformation properties of $u(\varphi)$ under the $G$ chiral group 
define a non--linear realization of the symmetry through the compensating
transformation $h(\varphi) \in SU(3)_V$~:
\begin{equation}
u(\varphi) \, \; \mapright{G} \, \;  g_R \, u(\varphi) \, 
h(\varphi)^{\dagger} \,
= \, h(\varphi) \, u(\varphi) \, g_L^{\dagger} \; \; \; \; \; \;  ,
\; \; \; \;  \; \; \; \; \; \; \; g_{L(R)} \,
\in SU(3)_{L(R)} \;  \; .
\end{equation}
Non--Goldstone bosons that belong to representations of $SU(3)$ (hence
transforming linearly under this group and nonlinearly under 
$SU(3)_L \otimes SU(3)_R$) can be included in the chiral 
Lagrangian following Ref.~\cite{cole}. We proceed in turn \footnote{We
do not consider light flavour or charmed singlets in the following. 
Their inclusion is
straightforward with our procedure.}~:
\begin{itemize}
\item[1/]{\bf Charmed mesons~:} \\
Charmed pseudoscalar mesons transform as triplets under $SU(3)$ and we
choose the representation~:
\begin{eqnarray} \label{eq:dosi}
D \, \equiv \, \left( \, \begin{array}{c}
                          \overline{D^0} \\
			  D^- \\
			  D_{\textsf{\footnotesize{s}}}^-
			  \end{array}
	     \, \right) 
	     \; \; \; & \; , \; \; & \; \; \; \; \; \; \; \; 	     
	     D \; \mapright{G} \; h(\varphi) \, D \; \; ,
\end{eqnarray}
and similarly for charmed resonances $D_R$~: vector ($D_{\mu}^V$), 
axial--vector ($D_{\mu}^A$) and scalar ($D^S$). We will introduce 
different masses for the various triplets of resonances. Within every
triplet we enforce the $SU(3)$ breaking of masses but we keep $SU(2)$
isospin symmetry.

\item[2/]{\bf Light resonances~:} \\
We are interested in resonances transforming as octets 
under $SU(3)$. Following Ref.~\cite{toni} and denoting by $R = V_{\mu}, 
A_{\mu}, S, ...$ these
octets, the non--linear realization of the chiral
group is given by~:
\begin{eqnarray}  \label{eq:rori}
R \; & \mapright{G} &  \; h(\varphi) \, R \, h(\varphi)^{\dagger} \; \; .
\end{eqnarray}
The flavour structure of $R$ is analogous to $\Pi$ in Eq.~(\ref{eq:uno}).	
To study the decays we are interested in we will need light vector
meson resonances that we introduce as Proca fields.
\end{itemize} 

We would like to establish, using the effective fields above, which
is the representation of the generating functional of QCD able to provide
matrix elements
of charged currents, responsible for semileptonic decays. To define
the relation of the effective action with QCD we may consider the 
effect of external sources $J$ that play the role of auxiliary variables 
\cite{jurg}.
The link between the underlying and the effective theory is given
by the Feynman path integral~:
\begin{equation}
e^{i \, \Gamma [ J ]} \, = \, N^{-1} \, \int \, 
[d \Pi] \, [d D_{(R)}] \, [d R] \, e^{i \, \int d^4 x \, {\cal L}_{eff}
[\, \Pi \, , \, \partial \Pi \, , \, D_{(R)} \, , \,  \partial D_{(R)} \, ,
\,  R \, , \,  \partial R \, ; \, 
J \, , \,  \partial J \, ]} \; \; , 
\end{equation}
where $N$ is the integral evaluated at $J = 0$. $\Gamma [J]$ on the
left--hand side is the generating functional of the Green functions
constructed with the operators of the underlying QCD, while the right--hand
side involves the effective field theory. The invariance of the 
generating functional under gauge transformations of the external 
sources implements the symmetry properties of the theory.
\par
Therefore
the weak interaction is introduced, similarly to the chiral gauge theory
framework, through external non--propagating fields. To realize the 
two weak $SU(2)_L$ doublets we now
couple the quarks $q = ( u, d, s, c)$ to $SU(4)$--valued hermitian external
fields $\tilde{\ell}_{\mu}$, $\tilde{r}_{\mu}$, $\tilde{s}$ and 
$\tilde{p}$~: 
\begin{equation}
{\cal L} \, = \, {\cal L}_{QCD}^{m=0} \, - \, m_c \, \overline{c} \, c \, 
+ \, \Frac{1}{2} \, \overline{q} \, \gamma^{\mu} \, \left[ \, 
\tilde {\ell}_{\mu}  \, (1-\gamma_5) \, + \, 
\tilde{r}_{\mu} \, 
(1+\gamma_5) \, \right]  q \, - \, \overline{q} \, 
( \, \tilde{s} \, - \, i \, \tilde{p}
\, \gamma_5 \,  ) \, q \; \; \; ,
\end{equation}
though we will only consider the left and right external sources that are
the ones needed to introduce the relevant interaction.
Note that in absence of external fields a mass term for the charmed quark $c$ 
remains.
\par
At the meson level the coupling of external sources requires a $SU(4)$ 
realization that embeds the two weak $SU(2)_L$ doublets into the effective
Lagrangian.
To proceed we construct a $4 \times 4$ matrix involving
light flavour and charmed pseudoscalars~:
\begin{eqnarray} \label{eq:cuatro}
\tilde{u}_R^{\dagger} \, = \, \left( \begin{array}{cc}
                               u(\varphi) & \Frac{i}{\sqrt{2} \, F_D}
			       u(\varphi) D \\ & \\
			      \Frac{i}{\sqrt{2} \, F_D} D^{\dagger} & F_D/F
			       \end{array}
		               \right) \; \; \; & , & \; \; \; \; \;
\tilde{u}_L \, = \, \left( \begin{array}{cc}
                               u(\varphi) &  \Frac{i}{\sqrt{2} \, F_D} D \\ 
			       & \\
			       \Frac{i}{\sqrt{2} \, F_D} D^{\dagger} u(\varphi)
			        & F_D/F
			       \end{array}
		               \right) \; \; , \nonumber \\
			       & &  \\
\widetilde{U} \, & = & \, \tilde{u}_R^{\dagger} \, \tilde{u}_L \; \; ,
\nonumber
\end{eqnarray}
and light flavour and charmed resonances~:
\begin{equation}
\widetilde{R} \, = \, \left( \begin{array}{cc}
                         R & D_R \\
			 D_R^{\dagger} & 0
			 \end{array}
                  \right) \; \; \; .
\end{equation}
However notice that, according with the transformation properties explained
above, light and charm flavoured pseudoscalar mesons enter with non--linear
and linear realizations, respectively. The role of the
$SU(4)$ realization in Eq.~(\ref{eq:cuatro}) is to help us to find out
the implementation of the external sources, in particular the charged 
current that relates the charm and light meson sector.
Therefore, by no means we are 
implying a seeming chiral realization with 4 flavours.
In Eq.~(\ref{eq:cuatro}) $F_D$ is the decay constant of charmed mesons 
(defined analogously to the $SU(3)$ octet decay constant $F$ that we 
identify with the decay constant of the pion).
\par
External chiral
sources, suitable for the introduction of weak interactions,
are coupled through the definition of covariant derivatives
on the relevant objects~:
\begin{eqnarray} \label{eq:prerl}
\Delta_{\mu} \, \widetilde{U} \, & = & \, \partial_{\mu} \widetilde{U} \, - \,
i \, \tilde{r}_{\mu} \, \widetilde{U} \, + \, i \, \widetilde{U} \, 
\tilde{\ell}_{\mu} \; \; \; , \nonumber \\
\nabla_{\mu} \widetilde{R} \, & = & \, \partial_{\mu} \widetilde{R} \, 
+ \, \left[ \, \widetilde{\Gamma}_{\mu} \, , \, \widetilde{R} \, \right]
\; \; , 
\end{eqnarray}
with
\begin{eqnarray}
\widetilde{\Gamma}_{\mu} \, & = & \, \Frac{1}{2} \, \left\{ \, 
\tilde{u}_R \, 
\left[ \partial_{\mu} - i \tilde{r}_{\mu} \right] \tilde{u}_R^{\dagger} \, + 
\, \tilde{u}_L \left[ \partial_{\mu}
- i \tilde{\ell}_{\mu} \right] \tilde{u}_L ^{\dagger} \, \right\} \; \; . 
\end{eqnarray}
The right-- ($\tilde{r}_{\mu}$) and left-- 
($\tilde{\ell}_{\mu}$) hand external fields are defined as an extension
of the $SU(3)$ case~:
\begin{eqnarray} \label{eq:rl}
\tilde{r}_{\mu} \, = \, \left( \begin{array}{cc}
                                r_{\mu} & 0 \\
				0 & \gamma_{\mu} 
				\end{array}
	                \right) \; \;  & \; , \; \; & 
			\; \;
\tilde{\ell}_{\mu} \, = \, \left( \begin{array}{cc}
                                \ell_{\mu} & \omega_{\mu} \\
				\omega_{\mu}^{\dagger} & \delta_{\mu} 
				\end{array}
	                \right)	 \; \; ,
\end{eqnarray}
and their transformation properties are chosen to give the covariant
character, under weak gauge transformations,
to derivatives in Eq.~(\ref{eq:prerl}). 
On the G/H coset space there are two Maurer--Cartan one--forms 
(left-- and right--chiral) related by parity~:
\begin{eqnarray} \label{eq:unoforms}
\textsf{l}_{\mu} & = & u ( \partial_{\mu} - i \ell_{\mu} ) u^{\dagger} \, = \,
\Gamma_{\mu} + (i/2) u_{\mu} \; \; , \nonumber \\
\textsf{r}_{\mu} & = & u^{\dagger} ( \partial_{\mu} - i r_{\mu} ) u \, = \,  
\Gamma_{\mu} - (i/2) u_{\mu} \; \; ,
\end{eqnarray}
which pullback to the space--time space
defines the axial vielbein $u_{\mu}$ and the
vectorial connection $\Gamma_{\mu}$. Stepping down to
$SU(3)$, the standard right- and 
left--handed currents are given by~:
\begin{eqnarray}
r_{\mu} \, & = & \, e \, Q \, \left( \, A_{\mu} \, - \, \tan \theta_W \, 
Z_{\mu} \, \right) \; \; , \nonumber \\
& & \\
\ell_{\mu} \, & = & \, 2 \, M_W \, \sqrt{\Frac{G_F}{\sqrt{2}}} \, 
                    \left( \begin{array}{ccc}
		   0 & V_{ud} W_{\mu}^{\dagger} & V_{us} W_{\mu}^{\dagger} \\
		   V_{ud}^* W_{\mu} & 0 & 0 \\
		   V_{us}^* W_{\mu} & 0 & 0 
		   \end{array}
		   \right) \;  \; \nonumber \\
		   & & \nonumber \\
	      & & \; + \, e \, Q \, A_{\mu} \, + \, e \, 
	      \left[ \Frac{1}{\sin 2 \theta_W} \, Q_L \, - 
	      \, Q \, \tan \theta_W \, \right] \, Z_{\mu} \; \; ,  \nonumber
\end{eqnarray}
with $Q = \frac{1}{3} \, diag \, ( 2, -1, -1)$ and $Q_L = diag \, (1,-1,-1)$.
The charmed mesons require a covariant
derivative on the $D_{(R)}$ triplets transforming under 
$SU(3)_L \otimes SU(3)_R$ as themselves~:
\begin{equation} \label{eq:nablad}
\nabla_{\mu} D_{(R)} \,  =  \, \left[ \, \partial_{\mu} \, + \, 
\Gamma_{\mu} \, + \, \Frac{i}{2} \, ( \gamma_{\mu} \, + \, \delta_{\mu} )  \,
\right] \, D_{(R)} \; \; \; , 
\end{equation}
where $\Gamma_{\mu}$ has been defined in Eq.~(\ref{eq:unoforms}) and the 
new chiral sources are~:
\begin{eqnarray}
\gamma_{\mu} \, & = & \, \Frac{2}{3} \, e \, \left[ \, A_{\mu} \, - \, 
                         \tan \theta_W \, Z_{\mu} \, \right] \; \; , 
			 \nonumber \\
		& &  \\
\delta_{\mu} \, & = & \, \Frac{2}{3} \, e \, A_{\mu} \, + \, e \, 
                    \left[ \, \Frac{1}{\sin 2 \theta_W} \, - \, 
		    \Frac{2}{3} \, \tan \theta_W \, \right] \, Z_{\mu} 
		    \; \; . \nonumber
\end{eqnarray}
Finally the left--handed field $\omega_{\mu}$ that drives the weak charged 
current interaction between the charmed and the light sector (as can be 
seen in Eq.~(\ref{eq:rl})) is given by~:
\begin{eqnarray} \label{eq:omechar}
\omega_{\mu} \, & = & \, 2 \, M_W \, \sqrt{\Frac{G_F}{\sqrt{2}}} \,
                    \left( \, \begin{array}{c}
		              0 \\
			      V_{cd}^* \\
			      V_{cs}^* 
			      \end{array}
		    \right) \, W_{\mu} \; \; ,
\end{eqnarray}
that under the chiral group $G$ transforms as
\begin{eqnarray}
\omega_{\mu} \; \; \mapright{G} \; \; g_L \; \omega_{\mu} \;  
& , \; \; \;  & \; \; \; \; \; \; \;    \; \; \; \; \; \; \; \; 
\;  \; \; \;  \; \; g_L \, \in \, SU(3)_L \; \; \; . 
\end{eqnarray}
We would like to emphasize that the electroweak gauge bosons introduced
here are not quantized, they behave as classical
fields and do not propagate.
\par
With these definitions we can provide the most general phenomenological
Lagrangian involving mesons with $u$, $d$, $s$ and $c$ quark content and
external fields implementing the weak chiral currents of the Standard
Model. However we are interested here in describing $D_{\ell 3}$ decays
that are brought about through charged current processes and we will 
limit ourselves to this case. Hence we design all
the relevant $SU(3)_L \otimes SU(3)_R$ gauge invariant operators. The
objects we need to carry on that construction are the effective field
realizations in Eqs.~(\ref{eq:uno},\ref{eq:dosi},\ref{eq:rori}), the
covariant derivative $\nabla_{\mu} D_{(R)}$ in Eq.~(\ref{eq:nablad}),
and the external charged source realization $\omega_{\mu}$ in 
Eq.~(\ref{eq:omechar}). All together with their transformation
properties under the gauge chiral group.
The resulting effective action is~:
\begin{eqnarray} \label{eq:seffo}
{\cal S}_{eff} \, & =  & \, \int \, d^4 x \, {\cal L}_{eff} \;   \; \; , 
\\  
{\cal L}_{eff} \, & = & \, {\cal L}_{\chi PT} \, + 
\, {\cal L}_{R \chi P T} \, + \, {\cal L}_{kin} \, + \, 
{\cal L}_D \, + \,{\cal L}_{D^S} \, + \, 
{\cal L}_{D^V} \, + \, {\cal L}_{D^A} \,  \, ,\nonumber
\end{eqnarray}
where ${\cal L}_{\chi PT}$ is the $SU(3)_L \otimes SU(3)_R$ 
chiral Lagrangian by
Gasser and Leutwyler \cite{jurg}, and ${\cal L}_{R \chi PT}$ is the
$SU(3)$ Lagrangian of the Resonance Chiral Theory \cite{toni}.
${\cal L}_{kin}$ collects all the kinetic and mass terms of charmed
mesons. It also
contributes to the interaction Lagrangian through the covariant derivatives. 
It reads~:
\begin{eqnarray}
{\cal L}_{kin} \, & = & \, ( \nabla^{\mu} D)^{\dagger} \, \nabla_{\mu} D \, 
 - \, D^{\dagger} \, {\cal M}_D \, D \; \; \nonumber \\
 & & \;  + \, ( \nabla^{\mu} D^S)^{\dagger} \, \nabla_{\mu} D^S \, 
 - \, (D^S)^{\dagger} \, {\cal M}_{D^S} \, D^S \; \; \\
 & & \; - \, \Frac{1}{2} \, ( D_{\mu \nu}^V )^{\dagger} \, (D^V)^{\mu \nu} \, +
 \, (D^V_{\mu})^{\dagger} \, {\cal M}_{D^V} \, (D^V)^{\mu} \; \; \nonumber \\
 & & \; - \, \Frac{1}{2} \, ( D_{\mu \nu}^A )^{\dagger} \,  (D^A)^{\mu \nu} \, +
 \, (D^A_{\mu})^{\dagger} \, {\cal M}_{D^A} \, (D^A)^{\mu} \; \; \nonumber \,
 \; \; ,
\end{eqnarray}
with $D^R_{\mu \nu} = \nabla_{\mu} D^R_{\nu} - \nabla_{\nu} D^R_{\mu}$, 
$R = V, A$, and 
the diagonal mass matrices ${\cal M}_{D^{(R)}}$ carry explicit 
$SU(3)$ breaking.
We give here in detail the remaining terms of Eq.~(\ref{eq:seffo})~:
\begin{itemize}
\item[-]{\bf Charmed pseudoscalars and light flavoured mesons} \\
\begin{eqnarray} \! \! \! \! \! \! \! \! \! 
{\cal L}_D \, & = & \, \Frac{F_D}{\sqrt{2}} \, 
                 \left[ \, \left( \nabla^{\mu} D \right)^{\dagger} \, u \, 
		 \omega_{\mu} \, + \, \omega_{\mu}^{\dagger} \, u^{\dagger} 
		 \, \nabla^{\mu} D \, \right]  
	       \, + \, i \, \Frac{\alpha_1 \, F}{2 \sqrt{2}} \, 
	      \left[ \, D^{\dagger} \, u^{\mu} \, u \, \omega_{\mu} \, - \, 
	      \omega_{\mu}^{\dagger} \, u^{\dagger} \, u^{\mu} \, D \,
	      \right] \nonumber \\
	      & & \, + \, i \, \Frac{\alpha_2 \, m_D^2}{4 \, F} \, 
	      \left[ \, D^{\dagger} \, V_{\mu} \, u \, \omega^{\mu} \, - \,
	      \omega_{\mu}^{\dagger} \, u \, V^{\mu} \, D \, \right] 
	      \, + \, i \, \beta_1 \, \left[ \, 
	      \left( \nabla^{\mu} D \right)^{\dagger} \, V_{\mu} \, D \, - \,
	      D^{\dagger} \, V_{\mu} \, \nabla^{\mu} D \, \right] \; \; .
	      \nonumber \\
	      & & 
\end{eqnarray} 
\item[-]{\bf Charmed scalars, charmed pseudoscalars and light flavoured 
mesons} \\
\begin{eqnarray} \! \! \! \! \! \! \! 
{\cal L}_{D^S} \, & = & \, i \, F_{D^S} \, 
                 \left[ \, \left( \nabla^{\mu} D^S \right)^{\dagger} \, u \, 
		 \omega_{\mu} \, - \, \omega_{\mu}^{\dagger} \, u^{\dagger} 
		 \, \nabla^{\mu} D^S \, \right] \, + \, 
		 \beta_2 \, \left[ \, D^{\dagger} \, u_{\mu} \, 
		 \nabla^{\mu} D^S \, + \, \left( \nabla^{\mu} D^S
		 \right)^{\dagger} \, u_{\mu} \, D \, \right] \nonumber \\
		 & & \, + \, \beta_3 \, \left[ \, \left( \nabla^{\mu} D 
		 \right)^{\dagger} \, u_{\mu} \, D^S \, + \, D^{S \, \dagger}
		 \, u_{\mu} \, \nabla^{\mu} D \,  \right] \, \,  .
\end{eqnarray}
\vspace*{0.1cm} 
\item[-]{\bf Charmed vectors, charmed pseudoscalars and light flavoured
mesons} \\
\begin{eqnarray} \! \! \! \! \! \! \! \! \! \! \! \! \! \! \!
{\cal L}_{D^V} \, & = & \, \Frac{F_{D^V} \, m_{D^V}}{2 \sqrt{2}} \, 
                   \left[ \, D_{\mu}^{\dagger} \, u \, \omega^{\mu} \, +
		   \, \omega_{\mu}^{\dagger} \, u^{\dagger} \, D^{\mu} \, 
		   \right] \, + \, i \, \beta_4 \, m_{D^V} \, 
		   \left[ \, D_{\mu}^{\dagger} \, u^{\mu} \, D \, - \, 
		   D^{\dagger} \, u_{\mu} \, D^{\mu} \, \right]  \nonumber \\
& & \, + \,  \Frac{\beta_{\varepsilon}}{2 \, m_D} \, 
\varepsilon_{\mu \nu \alpha \beta} \, \left[ \, D^{\dagger} \, V^{\mu \nu} \, 
\nabla^{\alpha} D^{\beta} \, + \, \left( \nabla^{\alpha} D^{\beta} 
\right)^{\dagger} \, V^{\mu \nu} \, D \, \right] \; .
\end{eqnarray}
\vspace*{0.1cm} 
\item[-]{\bf Charmed axial--vectors, charmed pseudoscalars and light 
flavoured mesons} \\
\begin{eqnarray} \! \! \! \! \! \! \! \! \! \! \! \! \! \! \!
{\cal L}_{D^A} \, & = & \, \Frac{F_{D^A} \, m_{D^A}}{2 \sqrt{2}} \,
		\left[ \, D_{\mu}^{A \, \dagger} \, u \, \omega^{\mu} \, + 
		\, \omega_{\mu}^{\dagger} \, u^{\dagger} \, D^{A \,\mu} \, 
		\right] \, + \, i \, \beta_5 \, m_{D^A} \, 
		\left[ \, D_{\mu}^{A \, \dagger} \, V^{\mu} \, D \, - \, 
		D^{\dagger} \, V^{\mu} \, D_{\mu}^{A} \, \right]  \; .
		\nonumber \\
		& & 
\end{eqnarray}
\end{itemize}
Here we have used 
$V_{\mu \nu} = \nabla_{\mu} V_{\nu} - \nabla_{\nu} V_{\mu}$, 
$\nabla_{\mu} V_{\nu} = \partial_{\mu} V_{\nu} + [ \Gamma_{\mu}, V_{\nu} ]$,
and
$m_{D^i}$, $i=S,V,A$ are typical mass scales for every $J^P$ introduced 
to define 
the dimensionless couplings $\alpha_i$, $\beta_i$ and 
$\beta_{\varepsilon}$. All together we have 12 a priori unknown
coefficients~: the decay constants $F_D$, $F_{D^S}$, $F_{D^V}$ and $F_{D^A}$,
and the couplings $\alpha_i$, $i=1,2$, $\beta_{\varepsilon}$ and 
$\beta_j$, $j=1,2,3,4,5$. Some information about masses is known and we 
may consider them as input in our study.
The interacting
effective Lagrangian ${\cal L}_{eff}$
provides a physical grounded parameterization of the 
$D \rightarrow P \ell^+ \nu_{\ell}$ and
$D \rightarrow V \ell^+ \nu_{\ell}$ processes without model--dependent
assumptions and hence it is a suitable basis for the analyses of 
experimental data. It is clear, though, that the number of unknown
couplings seems to undertone our task.
In the construction of ${\cal S}_{eff}$ we have exploited the rigorous
constraints that symmetries of the underlying QCD enforce on its
effective field theory.  However, symmetries give us the structure of
the operators but do not tell us anything about their coupling 
constants. In the next Section we will be back to this point.
\par
A thorough explanation of the features and
properties of the charm pieces of ${\cal L}_{eff}$ is now required. The
effective action of QCD in this energy region, as any effective field
theory, has an infinite number of pieces. We have collected only those
that contribute to $D_{\ell 3}$ processes with the fewer number of 
derivatives. This is so because, even if the included vertices can give
also contribution to $D_{\ell 4}$ processes, for example, many other
terms in the full effective action also contribute and should be taken 
into account. Though the chiral structure of the couplings might be 
suspect for the production of two or more non--soft light pseudoscalars,
it should be correct for the vertices under consideration where only
one light pseudoscalar is involved. This statement follows because,
on one side, fields are not observables and hence physics does not
depend on the field realization. In addition
hadron effective fields have very limited freedom in the structure of
their couplings and light pseudoscalars only can saturate Lorentz 
indices through derivatives. Moreover the requirement of chiral symmetry
not only
enforces the proper matching of the effective action at low energies.
Although chiral dynamics is often thought of as imposing constraints
only on low momentum processes, it also affects even the high energy
behaviour, a result worked out from analyticity \cite{mike}.
As a consequence, the structure of the couplings in our effective action
${\cal S}_{eff}$ is the most general one available for two-- and three--legs
vertices and, consequently, they should be able to describe both soft
and hard outgoing light pseudoscalar mesons. A similar situation happens
in the acknowledged Resonance Chiral Theory where, for example, the
$a_1(1260) \rightarrow \pi \gamma$ process is described along the same
lines we use in our effective action. We conclude that the structure
of the vertices in ${\cal S}_{eff}$ is the appropriate one to deal with
$D_{\ell 3}$ processes in all the energy range.
\par
Note that, contrarily to previous phenomenological Lagrangian approaches 
 in Ref.~\cite{Ca294,Casa106}, the construction of the effective
action of QCD that we have carried out does not rely in the heaviness of 
the charm quark but on the feature that non--Goldstone bosons belonging
to irreducible representations of $SU(N_F)$ can consistently be introduced
in an effective Lagrangian with the proper QCD symmetries 
\cite{steve,cole}. Sideways HQET
is an excellent perturbative framework to start with in 
the $B$ meson sector where inverse mass corrections are reasonably very 
small and provide the relevant breaking to the heavy quark symmetry
limit of QCD. Though rather massive it is not clear that this 
effective theory can be applied to the charm sector and, in any case, 
perturbative corrections would be much bigger, spoiling the convergence.

\section{Form factors in $D \rightarrow P \, \ell^+ \, \nu_{\ell}$ decays} 
\hspace*{0.5cm} $D_{\ell 3}$ processes with a pseudoscalar $P$ in the
final state are driven by a hadronic vector $H_{\mu}$ defined through
the amplitude of the decay~:
\begin{equation} \label{eq:mh}
M \,(D \rightarrow P \ell^+ \nu_{\ell}) \,  = \, - \, 
\Frac{G_F}{\sqrt{2}} \, V_{CKM} \,
\overline{u}_{\nu} \, \gamma^{\mu} \, (1-\gamma_5) \, v_{\ell} \, H_{\mu}
\; \; ,
\end{equation}
and that corresponds to the matrix element of the relevant vector
hadronic current driven by the $W_{\mu}$ 
field~: 
\begin{equation} \label{eq:left}
{\cal V}_{\mu} \, = \, 2 \, 
\Frac{\delta \, {\cal S}_{eff}}{\delta \, \omega^{\mu}} \, 
\biggl|_{J=0} \, \biggr. \; \; ,
\end{equation}
because only this current 
contributes to the processes under consideration. In Eq.~(\ref{eq:left})
$J$ is short for all external sources. Hence we obtain $H_{\mu}$ by 
differentiating the generating functional of our effective action. 
Its Lorentz decomposition is written out in terms of the
two independent hadron four--momenta in 
$D(p_D) \rightarrow P(p) \, \ell^+ \nu_{\ell}$ as~:
\begin{equation} \label{eq:curre}
H_{\mu} \, \doteq \, \,  \langle \, P(p) \, | \, {\cal V}_{\mu} \, e^{i \, 
{\cal S}_{eff} [J = 0]} \, | \, D (p_D) 
\, \rangle \, = \, 
f_{+}(q^2) \, (p_D + p)_{\mu} \, + \, f_{-}(q^2) \, (p_D - p)_{\mu} \; \; ,
\end{equation}
with $q^2 = (p_D - p)^2$, that introduces the two form factors associated
to the process. The $\exp (i {\cal S}_{eff}[J=0])$ term in
the definition of $H_{\mu}$ reminds us that the matrix element of the 
current has to be evaluated in presence of strong interactions. In terms 
of these form factors the spectrum of the semileptonic decay
is given by
\begin{eqnarray} \label{eq:spectre}
\Frac{d \, \Gamma (D \rightarrow P \ell^+ \nu_{\ell})}{d \, q^2} \, & = & 
\Frac{G_F^2 \, |V_{CKM}|^2}{384 \, \pi^3 \, m_D^3} \, 
\Frac{\sqrt{\lambda(q^2,m_D^2,m_P^2)}}{q^6} \, (q^2 - m_{\ell}^2)^2 \, \, 
\cdot \\
& & \left\{ \, |f_{+}(q^2)|^2 \, \left[ \, (2 q^2 + m_{\ell}^2) \, 
\lambda(q^2,m_D^2,m_P^2) \, + \, 3 \, m_{\ell}^2 \, (m_D^2 - m_P^2)^2 \, 
\right] \right. \nonumber \\
& & \, \, \, \, \,  \left. + \, 3 \, q^2 \, m_{\ell}^2 \, \left[ \, 2 \, 
Re(f_{+}(q^2) f_{-}^*(q^2)) \, (m_D^2 - m_P^2) \, + \, |f_{-}(q^2)|^2 \, q^2
\, \right] \, \right\} \; \; , \nonumber
\end{eqnarray}
where $\lambda(a,b,c) = (a+b-c)^2-4ab$ and, though not explicitly stated,
$f_{\pm}(q^2) \equiv f_{\pm}(q^2)[D,P]$.
When $m_{\ell} = 0$ the spectrum only depends on the $f_{+}(q^2)$ form factor
and therefore the dependence on $f_{-}(q^2)$ is suppressed, particularly 
for $\ell = e$.

\begin{figure}[tb]
\begin{center}
\hspace*{-0.5cm}
\includegraphics[angle=0,width=0.95\textwidth]{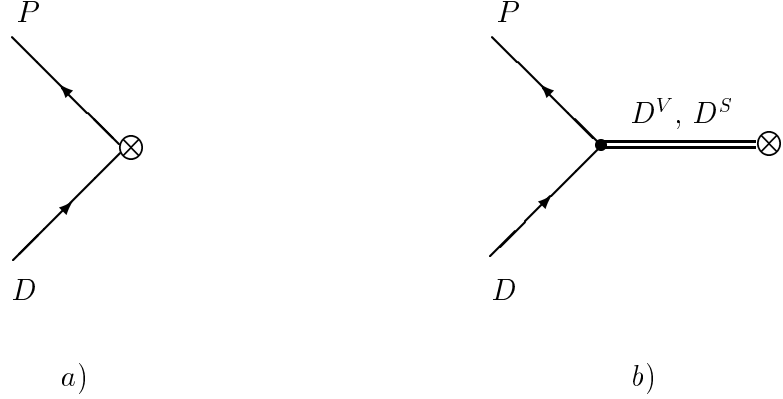}
\end{center}
\caption[]{\label{fig:uno} \it Tree-level contributions to $H_{\mu}$. 
The crossed circle indicates the external source insertion ${\cal V}_{\mu}$
and the black
dot is a strong interacting vertex. $D^V$ and $D^S$ are short for
charmed vector and scalar resonances, respectively.}
\end{figure}

\subsection{Form factors from the effective action in the $N_C \rightarrow
\infty$ limit}
\hspace*{0.5cm}
It has been widely emphasized \cite{Mano} that large number of colours
properties of QCD provide a guiding tool about basic features of the
strong interaction dynamics and, therefore, we intend to perform the evaluation
of the semileptonic form factors, defined above, at the leading order
in the $1/N_C$ expansion. To proceed we recall that the hadron matrix
element $H_{\mu}$ in Eq.~(\ref{eq:curre}) is related with the three--currents
Green function ${\cal G}_{\mu} \equiv \langle 0 | 
{\cal P}_D (x) {\cal P}_P(y) {\cal V}_{\mu}(z) | 0 \rangle$ where
${\cal P}_D(x)$ and ${\cal P}_P(x)$ are the pseudoscalar sources with
charm and light--quark quantum numbers, respectively, and 
${\cal V}_{\mu}(x)$ is the vector hadronic current in Eq.~(\ref{eq:left}).
The $1/N_C$ expansion gives precise information on the Green functions
of QCD currents \cite{HOO02}. In the $N_C \rightarrow \infty$ limit
the three--point function ${\cal G}_{\mu}$ is a sum of tree diagrams,
with free field propagators and local vertices. These diagrams are of 
two types~: in the first, one of the currents creates two mesons, each of
which is absorbed by the remaining currents (see Fig.~\ref{fig:uno}(a)), 
in the second each current creates one meson, and the three mesons combine
in a local vertex (see Fig.~\ref{fig:uno}(b)). Moreover one has to sum
over all the possible propagating mesons. At the next--to--leading order
in the $1/N_C$ expansion meson loops have to be taken into account.
\par
Coming back to our matrix element $H_{\mu}$ we see that the pseudoscalar
sources, creating the initial and final state mesons, are fixed 
and, in consequence, the $N_C \rightarrow \infty$ limit tells us that we
should consider the diagrams in Fig.~\ref{fig:uno} where, in (b) we must
sum over the infinite intermediate single resonances with 
${\cal V}_{\mu}$ quantum
numbers and with local couplings to $D$ and $P$ mesons. 
Within this approach and,
as in the Resonance Chiral Theory, we will assume that nearby resonances
provide most of the dynamics of the interaction; heavier resonance 
contributions being suppressed because their mass 
\footnote{In addition notice that only a single triplet of vector resonances
with the appropriate quantum numbers is known, and none of scalar resonances
\cite{pdg}.}.
Hence to proceed we evaluate the matrix element in 
Eq.~(\ref{eq:curre}) by approaching $\exp ( \, i \,  {\cal S}_{eff} \, ) \,
\sim \, 1 \,  + \,  i \, {\cal S}_{eff}$.  As we can 
see all the strong interaction, at this leading order, is reduced to the 
contribution in 
Fig.~\ref{fig:uno}(b) and it is mediated by charmed resonances. 
We obtain the following results~:
\begin{eqnarray}
M (D \rightarrow P \ell^+ \nu_{\ell}) \, & = &  \, - \, \Frac{G_F}{\sqrt{2}} \, 
\, \overline{u}_{\nu} \, \gamma^{\mu} \, (1-\gamma_5) \, v_{\ell} \, \, \cdot 
\, \\
& & \,  
 \, a(D,P) \, \cdot \,  \left[ \, f_{+}(q^2)[D,P] \, 
(p_D + p)_{\mu} \, + \, 
f_{-}(q^2)[D,P] \, (p_D - p)_{\mu} \, \right] \; \; , \nonumber
\end{eqnarray}
where $a(D,P)$ includes Clebsch-Gordan and Kobayashi--Maskawa couplings~:
\begin{eqnarray}
a(D^0, \pi^-) \, & = & \, - \, \sqrt{2} \; a(D^+, \pi^0) \, = \, 
a(D_{\textsf{\footnotesize{s}}}^+, K^0) \, =   \, V_{cd}^* \,
\; , \nonumber \\
& & \\
a(D^+, \overline{K^0}) \, & = & \, a(D^0, K^-) \, =   \, V_{cs}^* \; \; .
\nonumber
\end{eqnarray}
Form factors are given by~:
\begin{eqnarray} \label{eq:forma}
f_{+}(q^2)[D,P] \, & = & \, \Frac{1}{2} \, 
\left[ \, \Frac{F_D}{F} \, + \, \alpha_1 \, - \, \beta_4  \,  
\Frac{F_{D^V}}{F} \, \cdot \, \Frac{m_{D^V}^2}{q^2 \, - \, (M_{V}[D,P])^2} \, 
\right] \; \; , \nonumber \\
& & \\
f_{-}(q^2)[D,P] \, & = & \, \Frac{1}{2} \, 
\left[ \,  \Frac{F_D}{F} \, - \, \alpha_1 \, + \, 
2 \sqrt{2} \, \Frac{F_{D^S}}{F} \, \cdot \, 
\Frac{(\beta_3 \, - \, \beta_2) \, ( m_D^2 + m_P^2  - q^2 ) \, + \, 2 \, 
\beta_2 m_P^2}{q^2 \, - \, (M_{S}[D,P])^2} \, \right. \nonumber \\
& & \; \; \; \; \; \; \left. + \, \beta_4 \, \Frac{F_{D^V}}{F} \, \cdot \, 
\Frac{m_{D^V}^2}{(M_{V}[D,P])^2} \, \cdot \, 
\Frac{m_D^2 -m_P^2 - q^2 + (M_{V}[D,P])^2}{q^2 \, - \, (M_{V}[D,P])^2}  \, 
\right] \; \; . \nonumber
\end{eqnarray} 
The dependence on $D$ and $P$ in the form factors reduces to the masses
$m_D$, $m_P$ of the decaying and outgoing hadron, respectively, and 
$M_{V}[D,P]$, $M_{S}[D,P]$ appearing in the propagators in 
Eq.~(\ref{eq:forma}).
For the different channels we have~: 
\begin{eqnarray}
M_{V}[D^+,\pi^0]  \, & = & \, M_{V}[D^0, \pi^-] \, =
\, M_{V}[D_{\textsf{\footnotesize{s}}}^+, K^0 ] \, = \, m_{D^V} \, \; \; ,
 \nonumber \\
& & \nonumber \\
M_{V}[D^+ , \overline{K^0} ] \, & = & \, M_{V}[D^0 , K^- ]\, =
\, m_{D_{\textsf{\footnotesize{s}}}^V} \, \; \; , \nonumber \\
& &  \\
M_{S}[D^+ , \pi^0 ] \, & = & \, M_{S}[D^0 , \pi^- ] \, =
\, M_{S}[D_{\textsf{\footnotesize{s}}}^+ , K^0 ] \, = \, m_{D^S} \, \; \; ,
 \nonumber \\
& & \nonumber \\
M_{S}[D^+ , \overline{K^0} ] \, & = & \, M_{S}[D^0 , K^- ] \, =
\, m_{D_{\textsf{\footnotesize{s}}}^S} \, \; \; , \nonumber  
\end{eqnarray}
where the notation for the masses is self--explanatory. From the observed
spectrum of charmed mesons \cite{pdg}, $D^V$ would correspond to 
$D^*(2010)^{\pm}$, while $D_{\textsf{\footnotesize{s}}}^V$ 
corresponds to 
$D_{\textsf{\footnotesize{s}}}^*$. Scalar charmed resonances 
$D^S$ and $D_{\textsf{\footnotesize{s}}}^S$ 
still have not been observed.
\par
It is well known that $f_{-}(q^2)$ should vanish if $SU(4)_F$ 
symmetry is
exact due to the conservation of the vector current contributing to
the matrix element in Eq.~(\ref{eq:curre}). An inspection of our result
for $f_{-}(q^2)$ shows that to get that vanishing result it is not enough to 
enforce $m_D = m_P$ and $F_D = F$ and, therefore, the couplings of our
effective action are not independent from each other in the 
$SU(4)_F$ limit. This is not surprising
because the construction of our effective action ${\cal S}_{eff}$ was 
concerned with symmetry requirements from $SU(3)$ where the strong chiral 
realization lives and charmed mesons were introduced in a different
footing, as it should. It is more instructive, though, to leave this 
discussion to a later stage and we will come back to it.
\par
A next--to--leading evaluation in the $1/N_C$ expansion would provide,
typically, a $30 \%$ correction on our final results for $N_C = 3$,
although in other applications in resonance chiral theory these
are effectively smaller. In any case our approach would be good
enough for the analysis of present and foreseen experimental
results. The
computation of next--to--leading contributions is not feasible at the 
moment because we would need to 
consider the effective action at one loop, a non trivial task beyond the
scope of this work.

\subsection{QCD--ruled asymptotic behaviour of form factors}
\hspace*{0.5cm} The results that we have obtained for the $D_{\ell 3}$
form factors in Eq.~(\ref{eq:forma}) are a consequence of the
symmetry requirements enforced by QCD on our effective action
${\cal S}_{eff}$. 
 As commented at the end of Section 2, though, symmetries do not
constrain the coupling constants of ${\cal S}_{eff}$ and, consequently,
further insight is needed. To do so we remind the basic features of
effective field theory construction. Essentially this is an ongoing
procedure from the high energy scale to the energy region of interest 
where, in the stepping down, heavier degrees of freedom are integrated
out through an evolution process driven by both the renormalization group and
matching at the masses of heavier particles, when these decouple. We
do not explain in detail the construction \cite{George} but recall
two relevant conclusions for our work. First of all, when a heavy 
particle of mass $M$ is integrated out what results is a non--local
action. A later power expansion in $p/M$ ($p$ is a typical momentum
of the process) provides the final local non--renormalizable effective
action with derivative couplings, as our ${\cal S}_{eff}$. This already
tells us that the couplings in the effective action are going to be
suppressed by the masses of heavier degrees of freedom not present
in our action. The second conclusion of this procedure is that only
the short--distance information is incorporated into the coupling
constants of the effective Lagrangian \cite{George}. This is a 
powerful statement because, though we do not know how to evolve from
QCD down to the hadron level, it means that we can, and should, 
constrain the couplings according with the high energy behaviour of
the theory. And this indeed, we know, because asymptotic freedom 
provides a valid perturbative treatment of QCD at high energies.
\par
To proceed we will study the 
asymptotic behaviour ($q^2 \rightarrow \infty$) of form factors
of currents endowing, consequently, 
relations between the unknown couplings of ${\cal S}_{eff}$.
The restrictions on the semileptonic form factors involving
pseudoscalars imposed by their asymptotic behaviour ruled by 
QCD were already worked out time ago by Bourrely, Machet and
De Rafael \cite{eduard}.
\par
As we have said above $D_{\ell 3}$ decays with one pseudoscalar in the 
hadronic final state are driven by the vector current ${\cal V}_{\mu}$.
Then the form factors are related with the spectral functions 
associated to the vector two--point function \footnote{The relevant
flavour indices of the currents for every process should be understood.}~:
\begin{eqnarray} \label{eq:pimu}
\Pi_{\mu \nu} \, & = & \, i \, \int \, d^4 x \, e^{i \, q \, \cdot \, x} \,
\langle \, 0 \, | \, T \, ( {\cal V}_{\mu} (x) \, {\cal V}_{\nu}^{\dagger}
 (0) \, ) 
\, | \, 0 \, \rangle \, \, \nonumber \\
& = & \, - \, ( \, g_{\mu \nu} \, q^2 \, - \, q_{\mu} \, q_{\nu} \, ) \,
\Pi_1(q^2) \, + \, q_{\mu} \, q_{\nu} \, \Pi_0 (q^2) \, \, \; \,  ,
\end{eqnarray}
that are defined by~:
\begin{eqnarray} \label{eq:unitos}
 & & - \, ( \, g_{\mu \nu} \, q^2 \, - \, q_{\mu} \, q_{\nu} \, ) \,
 \mbox{Im} \Pi_1(q^2) \,  +   \, q_{\mu} \, q_{\nu} \, \mbox{Im} \Pi_0 (q^2) \,
 =   \! \! \\
 \! \! \! & & \, \Frac{1}{2} \sum_{\gamma} \, \int \, 
d \rho_{\gamma} \, 
(2 \pi)^4 \, \delta^{(4)}(q - p_{\gamma}) \, 
\langle \, 0 \, | \, {\cal V}_{\mu} (0) \, | \, \gamma \, \rangle \, 
\langle \, \gamma \, | \, {\cal V}_{\nu}^{\dagger} (0) \, | \, 0 \, \rangle \;
\; , \nonumber
\end{eqnarray}
where the summation is extended to all possible hadron states $\gamma$ with
appropriate quantum numbers, and the integration is carried on over the
allowed phase space of those states. In Eq.~(\ref{eq:pimu}) $\Pi_1$ 
corresponds to the contributions of $J^P = 1^-$ quantum numbers 
and $\Pi_0$ to those of $J^P = 0^+$. Between the infinite
number of intermediate contributions there is the one given by the
semileptonic matrix elements of $D \rightarrow P \ell^+ \nu_{\ell}$
given by Eq.~(\ref{eq:curre}) that we now write, not including the 
exponential of the effective action explicitly, as~:
\begin{equation} \label{eq:dos}
\langle \, 0 \, | \, {\cal V}_{\mu} (0) \, | \, D (p_D) \, \overline{P} (-p)
\, \rangle \, = \, \eta \, \left[ \left( \, p_D \, - \, 
\Frac{p_D \, \cdot \, q}{q^2} \, q \, \right)_{\mu} \, F_1(q^2) \, + \, 
\Frac{q_{\mu}}{q^2} \, F_0(q^2) \, \right]\; \; ,
\end{equation}
where $\eta$ is a Clebsch--Gordan coefficient and 
the new form factors, defined for convenience in the following
discussion, can be related with $f_{+}(q^2)$ and 
$f_{-}(q^2)$ through~:
\begin{eqnarray} \label{eq:newff}
F_1(q^2) \, & = & \, 2 \, f_{+}(q^2) \; \; , \nonumber \\
& &  \\
F_0(q^2) \, & = & \, ( \, m_D^2 \, - \, m_P^2 \, ) \, f_{+}(q^2) \, + \, 
q^2 \, f_{-}(q^2) \; \; . \nonumber
\end{eqnarray}
They correspond to $1^-$ and $0^+$ contributions, respectively.
Positivity of the spectral functions demands that every contribution
of the  $| \gamma \rangle $ intermediate states adds up and, therefore,
the two--pseudoscalar $| D \overline{P} \rangle$ state in the unitarity 
relation in Eq.~(\ref{eq:unitos}) is just one of the
infinite possible contributions to the spectral functions, to which it
provides a lower bound. Performing the phase space integration we
obtain~:
\begin{eqnarray} \label{eq:lowbo}
\mbox{Im} \, \Pi_1(q^2) \, & \ge &  \,
\Frac{\eta^2}{192 \, \pi} \, \sqrt{ \left( 1 - \Frac{Q_0^2}{q^2} \right)^{3
} \left(
1 - \Frac{Q_1^2}{q^2} \right)^{3}} \, |F_1(q^2)|^2 \, \, \theta(q^2 - Q_0^2)
\; \; , \nonumber \\
& & \\
\mbox{Im} \, \Pi_0(q^2) \, & \ge & \, 
\Frac{\eta^2}{16 \, \pi} \, \sqrt{\left( 1 - \Frac{Q_0^2}{q^2} \right) \left(
1 - \Frac{Q_1^2}{q^2} \right)} \, \Frac{|F_0(q^2)|^2}{q^4} \, \,
\theta(q^2 - Q_0^2) \; \; , \nonumber 
\end{eqnarray}
where $Q_0^2 = (m_D + m_P)^2$ and $Q_1^2 = (m_D - m_P)^2$.
\par
Perturbative QCD at leading order \cite{edu2} determines that
\begin{eqnarray} \label{eq:conligo}
\mbox{Im} \, \Pi_1(q^2) \, & \mapright{q^2 \rightarrow \infty} & \, 
\Frac{1}{4 \, \pi} \; \; , \nonumber \\
& & \\
\mbox{Im} \, \Pi_0(q^2) \, & \mapright{q^2 \rightarrow \infty} & \, 0 \; \; ,
\nonumber
\end{eqnarray}
and therefore, heuristically, one would expect that in the asymptotic regime
every one of the infinite positive contributions to the spectral function 
vanishes. This is clearly true for the $J=0$ spectral function and a 
reasonable guess for the $J=1$ vector function, expecting that the sum of the
infinite vanishing contributions gives a non--zero finite constant result.
Accordingly, from Eqs.~(\ref{eq:lowbo},\ref{eq:conligo}), we demand that the 
conditions~:
\begin{eqnarray} \label{eq:heffespe}
 F_1(q^2) \, & \mapright{q^2 \rightarrow \infty} & \, 0 \; \; ,  \nonumber \\
 & & \\
 F_0(q^2) \, & \mapright{q^2 \rightarrow \infty} & \, constant \; \; , \nonumber
\end{eqnarray}
are fulfilled.  In fact we could also choose that 
\footnote{{We comment later on the consequences of this stronger 
constraint.}}
$F_0(q^2) \rightarrow 0$ as $q^2 \rightarrow \infty$ but, while this is a
mandatory guess for $F_1(q^2)$, in the $J=0$ form factor this would
be a stronger condition that is not necessary, according with the heuristic
discussion above. From Eq.~(\ref{eq:lowbo}) we see that a constant
asymptotic behaviour is enough, and we attach to this softer assumption.
Nevertheless in both cases the constraints on the $f_{+}(q^2)$ and
$f_{-}(q^2)$ form factors are the same.
Using Eq.~(\ref{eq:newff}) we note that both $f_{+}(q^2)$
and $f_{-}(q^2)$ should vanish in the $q^2 \rightarrow \infty$ limit. Coming
back to Eq.~(\ref{eq:forma}) we get the following relations between the
couplings of the effective action ${\cal S}_{eff}$~:
\begin{eqnarray} \label{eq:asympcon}
& & \Frac{F_D}{F} \, + \, \alpha_1 \,  =  \, 0 \; \; \; , \nonumber \\
& & \\
& & 1 \, - \, \sqrt{2} \, \Frac{F_{D^S}}{F_D} \, (\beta_3 - \beta_2) \, -
\, \Frac{\beta_4}{2} \, \Frac{F_{D^V}}{F_D} \, 
\Frac{m_{D^V}^2}{(M_V[D,P])^2} \,  =  \, 0 \; \; \; . \nonumber
\end{eqnarray}
Carrying these relations to the expressions in Eq.~(\ref{eq:forma}) we get
the final parameterization of the form factors in 
$D \rightarrow P \ell^+ \nu_{\ell}$~:
\begin{eqnarray} \label{eq:final}
f_{+}(q^2)[D,P] \, & = &  \,  \, \, 
\Frac{\Omega[D,P]}{1 \, - \, \Frac{q^2}{(M_V[D,P])^2}} \; \; , \nonumber \\
& & \\
f_{-}(q^2)[D,P] \, & = & \, \Frac{m_P^2 \, - \, m_D^2}{(M_V[D,P])^2} \, \, 
f_{+}(q^2)[D,P] \, \, + \, \, \Frac{\Lambda[D,P]}{1 \, - \, 
\Frac{q^2}{(M_S[D,P])^2}} \; \; , \nonumber
\end{eqnarray}
where
\begin{eqnarray} \label{eq:final2}
\Omega[D,P] \, & = & \, \Frac{\beta_4}{2} \, \Frac{F_{D^V}}{F}  \, \, 
\Frac{m_{D^V}^2}{(M_V[D,P])^2} \, \; \; , \nonumber \\
& & \\
\Lambda[D,P] \, & = & \, \left( \, \Frac{F_D}{F} \, - \,  \,  
\Omega[D,P]  \, \right) \, 
\left( 1 \, - \,  \Frac{m_D^2}{(M_S[D,P])^2} \, - \, \Frac{m_P^2}{(M_S[D,P])^2}
  \, \right) \, \nonumber \\
& & \, - \, 2 \, \sqrt{2} \,\Frac{F_{D^S}}{F}  \, \beta_2 \, 
\Frac{m_P^2}{(M_S[D,P])^2} \; \; \; . \nonumber
\end{eqnarray}
These are our main results and Eq.~(\ref{eq:final}) shows the simplest 
parameterization of $f_{+}(q^2)$ and $f_{-}(q^2)$ consistent
with QCD constraints and saturation by resonances.
It is interesting to note that while our result for $f_{+}(q^2)$ coincides
with the phenomenological one--pole dominance approach shared by
other theoretical studies, $f_{-}(q^2)$ gets a two--pole structure
coming from vector and scalar resonances. This feature brings about into
the $0^+$ scalar $F_0(q^2)$ form factor, Eq.~(\ref{eq:newff}), the presence
of a local non--resonant contribution in addition to the one--pole
scalar meson resonance.
That local piece is induced by the $J^P=0^+$ time--like
polarization of the vector meson, through the cancellation of the vector 
resonance
pole introduced by the $f_{+}(q^2)$ term in $F_0(q^2)$.
\par
{ Our discussion above relies on the high--energy behaviour of the form 
factors in Eq.~(\ref{eq:heffespe}). As commented, strictly, QCD enforces
a constant (non necessarily vanishing) high--energy behaviour for 
$F_0(q^2)$. However, studies \cite{brolep} that assume factorization
at high $q^2$ and some common lore physics intuition would demand the 
stronger $F_0(q^2) \, \mapright{q^2 \rightarrow \infty} \, 0$ condition. Hence
$f_{-}(q^2)|_{q^2 \rightarrow \infty}$ would vanish at least as 
$1/q^4$ requiring, consequently, a pure double pole structure. This would
enforce an extra condition on the couplings of the effective lagrangian,
namely, $\Lambda[D,P] = \Frac{m_D^2 - m_P^2}{(M_S[D,P])^2} \, \Omega[D,P]$.
We call $\Lambda[D,P]_{FACT}$ this value for $\Lambda[D,P]$.
The experimental measurement of the $f_{-}(0)$ would provide, in consequence,
a relevant
information on the QCD structure of the form factors.}
\par
As commented above, in the $SU(4)_F$ limit $f_{-}(q^2)$ should
vanish. We observe that this constraint determines relations between
the couplings that are only valid in that limit. Hence we get that
$\Lambda[D,P] |_{SU(4)} \, = \, 0$, that provides a relation between
the couplings in this limit.
However it is clear that 
$N_F = 4$ flavour symmetry is badly broken and therefore this 
condition should not be taken seriously. 
\par
Pion pole dominance and $SU(2)_L \otimes SU(2)_R$ current algebra
provide the Callan--Treiman relation between
the $K_{\ell 3}$ from factors and the decay constant of kaon $F_{K}$
that drives $K_{\ell 2}$ decays 
\cite{callan}~: 
\mbox{$f_{+}^{K\pi}(m_K^2) + f_{-}^{K\pi}(m_K^2) = F_K/F$},
in the vanishing pion mass limit. In our case a direct 
evaluation, 
from Eqs.~(\ref{eq:final},\ref{eq:final2}), gives~:
\begin{eqnarray} \label{eq:callan}
\Frac{F_0(m_D^2-m_P^2)}{m_D^2-m_P^2} \; \bigg\vert_{[D,P]}  \; & = 
& \; f_{+}(m_D^2-m_P^2) \, 
+ \, f_{-}(m_D^2-m_P^2) \; \vert_{[D,P]}   \\
  & = & \; \Frac{F_D}{F} \,  - \, 
2 \, \Frac{m_P^2}{(M_S[D,P])^2-m_D^2+m_P^2} \, \left[ \, 
\Frac{F_D}{F} \, + \, \sqrt{2} \Frac{F_{D^S}}{F} \, \beta_2 \, 
-  \, \Omega[D,P] \, \right] \,  . \nonumber
\end{eqnarray}
Note that the evaluation point,
$q^2 = m_D^2 - m_P^2$, is outside the physical region. 
 In the $SU(3)$ chiral limit $m_P=0$ and we have
$f_{+}(m_D^2)+f_{-}(m_D^2) \, |_{[D,P]}^{\chi} = F_D/F$ as the Callan--Treiman
relation endows when applied to the four flavour case. Although the
$m_P = 0$, $P=\pi,K$, limit in Eq.~(\ref{eq:callan}) seems affordable,
nothing can be said about the size of the correction because our lack
of knowledge on the couplings. However notice that a strong cancellation
in the denominator of that term~: $(M_S[D,P])^2-m_D^2+m_P^2$, if charmed
scalar resonances are near, could provide a sizeable contribution.

\section{Phenomenology of $D \rightarrow P \ell^+ \nu_{\ell}$}
\hspace*{0.5cm} As we said in the Section 1 experiment FOCUS (E831) at
Fermilab is foreseen to provide, in the near future, a thorough
study of semileptonic form factors of charmed mesons. Until present
several observables have been measured with rather good accuracy 
\cite{oldies,old4,old3,old2,cleito1,cleito2,cleito3} but
an exhaustive study of the $q^2$ behaviour of form factors, even the
dominant $f_{+}(q^2)$, is still missing. 
\par
The exclusive channels studied up to now are the Cabibbo--favoured
$D^+ \rightarrow \overline{K^0} \ell^+ \nu_{\ell}$,
$D^0 \rightarrow K^- \mu^+ \nu_{\mu}$, and the Cabibbo--suppressed
$D^+ \rightarrow \pi^0 \ell^+ \nu_{\ell}$ and 
$D^0 \rightarrow \pi^- e^+ \nu_{e}$,
which branching ratios are measured reasonably well. The study of the
$q^2$--structure of their form factors, however, is much poor. 
Notwithstanding, experiment E687 has published reasonable spectra 
in $D^0 \rightarrow K^- \mu^+ \nu_{\mu}$ \cite{old3} though we have
been advised \footnote{Private communication received from Will
Johns.} that they are not corrected for background, 
resolution and acceptance effects and, consequently, should not be
used to analyse theoretical form factors.
\par
From
an experimental point of view, the data is usually 
fitted to one--pole form factors~:
\begin{equation} \label{eq:experff}
f_{\pm}(q^2) \, \,  = \, \, \Frac{f_{\pm}(0)}{1 \, - \, 
\Frac{q^2}{m_{\pm}^2}} \; \; ,
\end{equation}
though due to the $m_{\ell}$--suppression pointed out in our discussion
related with Eq.~(\ref{eq:spectre}) the $f_{-}(q^2)$ is very much
unknown. Other parameterizations for $f_{+}(q^2)$ are also possible. 
In particular, and due to pioneering modelizations \cite{isgo}, the
exponential behaviour 
$f_{+}(q^2) \, = \, f_{+}(0) \, \exp ( \alpha \, q^2)$ has also been
fitted to data. Nevertheless in the available range of energies it is not
possible to distinguish both parameterizations. However from experiment
one gets $\alpha = (0.29 \pm 0.7) \, \mbox{GeV}^{-2} \,   > \, 0$ 
\cite{cleito1} and, therefore, the asymptotic behaviour of
this last parameterization is disastrous according with our discussion
in Section 3. Surely the exponential form factor is not consistent
with QCD. Moreover, notice that a one--pole form factor only for $f_{-}(q^2)$,
as in Eq.~(\ref{eq:experff}), is not allowed (unless 
$f_{+}(q^2) = 0$) because $F_0(q^2)$ in Eq.~(\ref{eq:newff}) would 
drive a $J^P = 1^-$ transition, through the pole of the vector resonance,
that is forbidden for that form factor.

\begin{table}
\begin{center}
\begin{tabular}{|c|c|c|} 
\hline
& & \\
\multicolumn{1}{|c|}{Experiment} &
\multicolumn{1}{|c|}{$| f_{+}(0) | $} & 
\multicolumn{1}{|c|}{$m_{+} \mbox{(GeV)} $} \\
& & \\
\hline
\hline
& &  \\
E687 \protect{\cite{old3}}&  $0.71 \pm 0.04$ &  $1.87 \pm 0.13$ \\
& &  \\
\hline
& &  \\
CLEO \protect{\cite{cleito1}} & 
$0.77 \pm 0.04$ & $ 2.00 \pm 0.22$ \\
& &  \\
\hline
\end{tabular} 
\caption{\it Experimental values for $|f_{+}(0)|$ and $m_{+}$ from
$D^0  \rightarrow K^- \ell^+ \nu_{\ell}$ decays.}
\end{center}
\end{table}

Hence in $f_{+}(q^2)$ there are two parameters to fit~: 
$f_{+}(0)$ and the pole mass $m_{+}$. Experimental figures are 
collected in Table 1. From our result in Eq.~(\ref{eq:final}) we see that
\begin{equation} \label{eq:atcero}
f_{+}(0)[D,P] \, \, = \, \, \Omega[D,P]  \; \; .  
\end{equation}
Hence, from experiment, $| \Omega[D^0,K^-] | \simeq 0.75$  in excellent
agreement with sum rules expectations \cite{SUMO}.
On the other side the obtained values of $m_{+}$
are consistent with the experimental value of 
$m_{D_{\textsf{\footnotesize{s}}}^V} = 
m_{D_{\textsf{\footnotesize{s}}}^*} = 2.1124 \pm 0.0007 \, \mbox{GeV}$
that is the one appearing in our form factor.
\par
The study of the ratio 
$Br(D^0 \rightarrow \pi^- \ell^+ \nu_{\ell})/ Br(D^0 \rightarrow K^- \ell^+
\nu_{\ell})$ 
provides information over the difference between $f_{+}(q^2)$ form factors
with $K$ or $ \pi$ in the final state. Experimental figures are given 
in Table 2. From our results we predict~:
\begin{equation}
\Frac{|f_{+}(0)[D^0,\pi^-]|}{|f_{+}(0)[D^0,K^-]|} \, \, = \, 
\, \Frac{m_{D_{\textsf{\footnotesize{s}}}^V}^2}{m_{D^V}^2} \, \, 
\simeq \, 1.05  \; \; \; ,
\end{equation}
if we take, from Ref.~\cite{pdg}, 
$m_{D_{\textsf{\footnotesize{s}}}^V} = m_{D_{\textsf{\footnotesize{s}}}^*}$ and
$m_{D^V} = m_{D^*(2010)^{\pm}}$. 
\par
The structure provided by our approach for $f_{-}(q^2)$ in 
Eq.~(\ref{eq:final}) is much more complex. We have a two--pole structure
that it would be very much interesting to explore phenomenologically.
Unfortunately,
to our knowledge, the only known experimental result on $f_{-}(q^2)$ is 
provided by the E687 Collaboration \cite{old3} that give 
\begin{equation}
\Frac{f_{-}(0)[D^0,K^-]}{f_{+}(0)[D^0,K^-]} \, \, = \, \, - \, 1.3 
\pm_{3.4}^{3.6}
\; \; ,
\end{equation}
still compatible with zero, to compare with our result~:
\begin{equation} \label{eq:raton}
\Frac{f_{-}(0)[D^0,K^-]}{f_{+}(0)[D^0,K^-]} \, \, = \, \, - \, 
\Frac{m_{D^0}^2 \, - \, m_{K^-}^2}{m_{D_{\textsf{\footnotesize{s}}}^V}^2} \, + \, 
\Frac{\Lambda[D^0,K^-]}{\Omega[D^0,K^-]} \; \, .
\end{equation}
In our prediction the first term gives 
$(m_{K^-}^2 - m_{D^0}^2)/m_{D_{\textsf{\footnotesize{s}}}^V}^2 \simeq -0.72$,
 agreeing in sign and size with the central value
in the experimental determination.
{If we take $\Lambda[D^0,K^-]_{FACT}$ we would have a prediction 
for the ratio
in Eq.~(\ref{eq:raton}) in terms of masses of resonances and pseudoscalars.
Unfortunately the unknown scalar charmed meson mass is also involved.}  
Although we know very little about
$\Lambda[D,K]$ from the phenomenology, determinations of $f_{\pm}(0)$ within
a sum rules approach provide information on $\Lambda$. 
With the results
of Ref.~\cite{SUMO} we find $\Lambda[D^0,K^-] = -0.05 \pm 0.11$ and
$\Lambda[D^0,\pi^-] = -0.03 \pm 0.12$, hence compatible with zero. Accordingly
the ratio in Eq.~(\ref{eq:raton}) is very well approximated by the first
term only and, in addition, we can conclude that the contribution of the
scalar resonances to $f_{-}(q^2)$ in Eq.~(\ref{eq:final}) should be tiny.
{ Moreover notice that the sum rules predictions are at odds with 
$\Lambda[D,P]_{FACT}$ unless the lightest scalar charmed resonance has a very
large mass.}

\begin{table}
\begin{center}
\begin{tabular}{|c|c|} 
\hline
& \\
\multicolumn{1}{|c|}{Experiment} &
\multicolumn{1}{|c|}{$\Frac{|f_{+}(0)[D^0,\pi^-]|}{|f_{+}(0)[D^0,K^-]|}$} \\
& \\
\hline
\hline
&   \\
E687 \protect{\cite{old4}}&  $1.00 \pm 0.12$ \\
&  \\
\hline
&  \\
CLEO \protect{\cite{cleito3}} &  $ 1.01 \pm 0.21$ \\
&   \\
\hline
\end{tabular} 
\caption{\it Experimental values for the ratio $|f_{+}(0)[D^0,\pi^-]|/
|f_{+}(0)[D^0,K^-]|$. It has been used that $(|V_{cd}|/|V_{cs}|)^2 \, = 
\, 0.051 \pm 0.001$. }
\end{center}
\end{table}

In conclusion much more work is needed on the experimental side to be
able to compare our results with the phenomenology. The spectrum of the
semileptonic decays of charmed mesons should be measured with good 
accuracy in order we can confirm the structure of $f_{+}(q^2)$ and
find out if the two--pole peculiar feature of the QCD and saturation 
by resonances driven $f_{-}(q^2)$ is confirmed. With these 
analyses we could give a serious step forward in the determination 
and comprehension of the
effective action of QCD in the charm energy region.

\section{Other semileptonic decays}
\hspace*{0.5cm} The effective action ${\cal S}_{eff}$ in Eq.~(\ref{eq:seffo})
allows us to evaluate all semileptonic $D_{\ell 3}$ and $D_{\ell 4}$ decays.
A thorough phenomenological study of them would provide a good knowledge
on the couplings of the operators in ${\cal L}_{eff}$ that determine
their strength. In this first paper we have addressed the study of the
simplest processes $D \rightarrow P \ell^+ \nu_{\ell}$ with the conclusions
pointed out in Sections 3 and 4. We stress here the interrelation between
the couplings and other processes.

\begin{table}
\begin{center}
\begin{tabular}{|c|c|} 
\hline
& \\
\multicolumn{1}{|c|}{Processes} &
\multicolumn{1}{|c|}{Couplings} \\
& \\
\hline
\hline
& \\
$\; D \, \rightarrow \, P \,  \ell^+ \nu_{\ell} \; $ &  $F_D$, $ \, \alpha_1 \, $, 
$\, \beta_4 \, F_{D^V} \, $, $\,  \beta_2 \, F_{D^S} \, $, 
$ \, \beta_3 \, F_{D^S}$   \\
& \\
\hline
& \\
$\, D \, \rightarrow  \, P \,  P \,  \ell^+ \nu_{\ell} \,$ & $F_D \, $, 
$\, \alpha_1 \, $,
$\, \beta_4 \, F_{D^V}$  \\
&   \\
\hline
& \\
$\; D \, \rightarrow  \, V \, \ell^+ \nu_{\ell} \;$ & 
$\alpha_2 \, $, $\, \beta_1 \, F_D \, $, 
$\, \beta_{\varepsilon} \, F_{D^V} \, $, 
$\, \beta_5 \, F_{D^A} \, $ \\
& \\
\hline
\end{tabular} 
\caption{\it Couplings or combinations of couplings from ${\cal L}_{eff}$
appearing in the form factors of semileptonic decays of charmed mesons. As in
the main text $P$ is short for a light pseudoscalar meson and $V$ is short
for a light vector meson.}
\end{center}
\end{table}

Decay constants of mesons parameterize the transition from the meson to the 
hadronic vacuum.
While there is a reasonably good knowledge on the $D$ decay constant
$F_D$ \cite{pdg} from $D_{\ell 2}$ decays, the phenomenological determination
of the decay constants
of resonances $F_{D^S}$, $F_{D^V}$ and $F_{D^A}$ involves electroweak 
decays (such as $D_R \rightarrow \ell^+ 
\nu_{\ell}$,...) that are tiny against the strong dominating processes.
Therefore their experimental evaluation is out of question. 
In addition $\beta_{\varepsilon}$, $\beta_1$ and $\beta_5$ only
appear in off--shell strong vertices. The strong couplings
$\beta_2$, $\beta_3$ and $\beta_4$in ${\cal L}_{eff}$ could be determined 
from on--shell strong processes.   Although the first two involve still 
unobserved scalar charmed resonances, the $\beta_4$  coupling,
that drives $D^* \rightarrow D \pi$, can be obtained through the recent
observation of this decay \cite{desept}. From this width we get
$|\beta_4| \, = \, 0.58 \pm 0.07$. Notice that $\beta_4$ is involved in
the determination of $f_{+}(0)$ (see Eqs.~(\ref{eq:final2},\ref{eq:atcero}))
however we do not know the value of the decay constant of vector charmed
resonances, $F_{D^V}$, and consequently we cannot predict $f_{+}(0)$ in
a model--independent way. Reversely, using its experimental value we can
determine $|F_{D^V}| \sim 240 \, \mbox{MeV}$.

\par
The role of the phenomenology of semileptonic processes to 
get information on these couplings is relevant. In these decays we usually
have amplitudes that involve one coupling, like the vertex in Fig. 1(a), or
the product of two couplings, as the two connected vertices in Fig. 1(b).
A close look to the ${\cal L}_D$, ${\cal L}_{D^S}$,
${\cal L}_{D^V}$ and ${\cal L}_{D^A}$ Lagrangians shows the couplings
relevant for the different processes. We collect them in Table 3. Notice
that $D \rightarrow V \ell^+ \nu_{\ell}$ processes also contribute to 
$D_{\ell 4}$ decays through a strong conversion $V \rightarrow P P$ driven
by ${\cal L}_{R \chi PT}$ in Eq.~(\ref{eq:seffo}) which couplings are rather
well known.  
\par
The foreseen good prospects on the experimental side for the near future,
together with the QCD constraints from the dynamical behaviour in the
asymptotic limit (not taken into account when writing Table 3), that also 
should extend properly to $D \rightarrow V \ell^+ \nu_{\ell}$ and 
$D_{\ell 4}$ processes, would be able to determine reasonably well the 
effective action of QCD in this energy regime.

\section{Comparison with the heavy quark mass expansion}
\hspace*{0.5cm}
An alternative approach based also in a phenomenological Lagrangian
that tries to implement both HQET \cite{hqs} and chiral symmetry 
\cite{jurg} has been employed during the last years \cite{Ca294,Casa106} in the
study of $\mbox{heavy} \rightarrow \mbox{light}$ semileptonic processes.
This is a rigorous and systematic procedure that deals with the
construction of an 
effective action of QCD through the constraints of Heavy Quark and 
Chiral symmetries and that inherites from HQET the perturbative expansion
in inverse powers of the heavy quark mass, typically $m_q/M_Q$ and
$\Lambda_{QCD}/M_Q$ where $m_q$ and $M_Q$ are the masses of light and heavy
quarks, respectively.
\par
This feature brings several consequences. On one side the fast convergence
of the perturbative expansion in the study of $B$ meson decays, because of
the high mass of the $b$ quark, does not apply so clearly in the case
of $D$ meson decays. Moreover although the effective action is very
simple in the $M_Q \rightarrow \infty$ limit, where the nice property of
relating $B$ and $D$ processes arises, it becomes rather cumbersome
when next--to--leading terms in the mass expansion are included, 
consequently loosing predictability, unless
some modelization hypotheses are assumed \cite{IT195}. On the other side,
heavy--quark symmetry relations are useful if the recoiling light
constituents can only probe distances that are large compared with 
$1/M_Q$. This condition is equivalent to the statement 
$(v\cdot v' -1) \ll M_Q/\Lambda_{QCD}$ or 
$q^2 \simeq q_{max}^2 = (m_D - m_P)^2$ in $D \rightarrow P \ell^+ \nu_{\ell}$
processes, where $v$ and $v'$ are the four--velocities of the initial and
final hadrons. Hence in this framework one evaluates 
$f_{\pm}(q_{max}^2)$, a particular analytic
continuation for the form factors (usually a monopole structure given
by vector meson dominance) is 
assumed and, in consequence, a prediction for $f_{\pm}(0)$ is given. It is
necessary to emphasize that the prediction of the form factors, given
by the heavy quark mass expansion, at $q^2 \neq q^2_{max}$ includes 
input from outside the perturbative treatment.
\par
The effective theory framework that we propose in this article, on the
other hand, relies on well known aspects of the underlying QCD theory.
We skip the heavy quark mass expansion by applying the well--known
procedure of constructing a phenomenological Lagrangian \cite{cole}
on the basis of chiral symmetry (for the light flavours) and considering
the charm flavoured mesons as matter fields in specific $SU(3)$
representations that provide the interaction.  The phenomenological 
Lagrangian acquires specific features of QCD by imposing the high--energy
behaviour on the form factors, procedure that constrains the couplings.
This is an essential step in the construction of the effective action
in order to improve our Lagrangian with another model--independent tool
that facilitates the matching at higher energies. All this
methodology is analogous to the one used in the Resonance Chiral Theory.  
In addition the dynamical structure of the form factors does not rely
in assumed analytic continuations but on the prediction of QCD in the
limit of large number of colours ($N_C \rightarrow \infty$). As emphasized
above this limit establishes the role of single resonances in the
Green functions and, consequently, in our form factors. 
Notice that the procedure we are presenting may be extended systematically
by including next--to--leading corrections in the $1/N_C$ expansion though,
as in the heavy quark mass expansion, one needs to perform the
construction of the action at one--loop level.
\par
A complete comparison between our results for the semileptonic form
factors $f_{\pm}(q^2)$ and those of the heavy mass expansion at
leading order (that
we take from Ref.~\cite{Casa106} for definiteness) is not feasible
because the different input included in their construction.
The main difference arises from the high energy constraints on our 
effective action. These have no clear meaning in the heavy quark mass
expansion when one perturbates around the heavy quark mass. On the 
other side our results for the form factors (\ref{eq:final}) 
include the contributions of scalar resonances that have not been taken
into account in the heavy mass expansion approach. However it is easy
to see that, switching off these scalar contributions and performing
a heavy meson mass expansion on our results in Eq.~(\ref{eq:final}), we
recover the features of the heavy quark mass expansion results 
in Ref.~\cite{Casa106}~:
\begin{eqnarray}
f_{+}(q^2) + f_{-}(q^2) \, & \simeq & \, 2 \, \Frac{\Delta}{M_V} \,
f_{+}(q^2) \; , \nonumber \\
f_{+}(q^2) - f_{-}(q^2) \, & \simeq & \, 2 \, f_{+}(q^2) \; ,
\end{eqnarray}
where $\Delta = M_V - m_D$. Moreover, in this limit, our results
for $f_{+}(q^2)$ coincide with those of that reference 
provided that $|\beta_4 F_{D^V}/2| = |g F_D|$,
where $g$ drives the $D^* \rightarrow D \pi$ decay in HQET.
 From this last process
we see that $|\beta_4| = |g|$ and, in addition, with our definition of 
the decay constants the heavy quark spin symmetry demands that 
$F_{D^V} = 2 F_D$. Hence the consistency of our prediction for the form
factors $f_{\pm}(q^2)$ with the heavy quark mass limit is exact. However
we stress that our results include the mass corrections to that limit.
\par
In Section 5 we got that $|F_{D^V}| \sim 240 \, \mbox{MeV}$. The heavy 
quark spin symmetry demands that $F_D = F_{D^V}/2 \sim 120 \, \mbox{MeV}$ and,
experimentally, the value of this decay constant is still rather uncertain,
$F_D = 212 \pm^{139}_{109} \, \mbox{MeV}$ \cite{pdg}. Notice however that,
as emphasized in Ref.~\cite{Neube}, the spin--symmetry--breaking effects in
the charmed sector could be as large as $50 \%$.

\section{Conclusions}
\hspace*{0.5cm} The study of form factors of QCD currents provides
all--important information on the relevant effective action of the
underlying theory. Semileptonic decays of mesons are the main tool
to analyse charged currents and, while $B$ and $K$ decays have
received very much attention, $D$ decays, due to their position
in the energy spectra, lack a definite and sounded framework where
to root this task.
\par
We have proposed a model--independent scheme that relies in the
use of phenomenological Lagrangians generated through the symmetries
of QCD  and the dynamics of its $N_C \rightarrow \infty$ limit.
 In this scheme the three lightest flavours are introduced 
following the guide of chiral symmetry while charmed mesons appear
as matter fields, following Refs.~\cite{steve,cole}. The procedure
is analogous to the construction of the Resonance Chiral Theory 
\cite{toni}. Hence we arrive to an effective field theory where
the structure of the operators is driven by the symmetries and 
their couplings are unknown. In addition, the QCD--ruled asymptotic
behaviour of form factors imposes several constraints on those
couplings.  In this framework, we have computed the form factors
in the semileptonic $D \rightarrow P \ell^+ \nu_{\ell}$ processes
at leading order in the $1/N_C$ expansion and we end with the 
parameterizations in Eq.~(\ref{eq:final}) that are our main result.
It is necessary to emphasize that this approach is different from 
the one followed in Refs.~\cite{Ca294,Casa106} that relies
in the heaviness of the charmed quark while here this consideration,
with its possible misconceptions in the charmed case,
does not appear.  Moreover we do not need to assume a particular
structure for the analytic continuation of the form factors because
we rely in the dynamics driven by the $N_C \rightarrow \infty$ limit
of QCD.
\par
The experimental situation in $D \rightarrow P \ell^+ \nu_{\ell}$ is
rather poor though it is foreseen to upgrade in the near future.
While our result for $f_{+}(q^2)$
is consistent with experimental analyses,  we 
would consider very much interesting that, through 
$D \rightarrow P \mu^+ \nu_{\mu}$ processes, something could be said 
on the $f_{-}(q^2)$ form factor, where we have concluded that a two--pole
structure is predicted in our framework. The parameterization we propose,
to analyse the experimental data, is then~:
\begin{eqnarray}
f_{+}(q^2) \, \, & = & \, \, \Frac{a_V}{1 \, - \, \Frac{q^2}{M_V^2}} \; \; ,
\nonumber \\
& & \\
f_{-}(q^2) \, \, & = & \, \, \Frac{b_V}{1 \, - \, \Frac{q^2}{M_V^2}} \, + \, 
\Frac{b_S}{1 \, - \, \Frac{q^2}{M_S^2}} \; \; \; . \nonumber
\end{eqnarray}
At present $a_V$, $b_V$ (that is proportional to $a_V$ according to
our prediction in Eq.~(\ref{eq:final})) and $M_V$ are rather well known.
However nothing can be said
about the size of $b_S$ nor $M_S$ (scalar resonances with charm have 
not been observed) and, consequently, this should be an important task
for future research in this field. 
If one uses the definition of form factors in Eq.~(\ref{eq:newff})
instead, $F_0(q^2)$ should show, in addition to the one--pole structure
induced by the scalar resonances, a non--negligible local piece 
acting as a background.
Once all this observables are measured
we will be able to constrain the effective action by determining better
the strength of its operators.
\par
A complementary study of the form factors in 
$D \rightarrow V \ell^+ \nu_{\ell}$
within the effective action of QCD proposed in this paper is under way
\cite{noi}.
\par
Finally we have shown that, while it is not possible to apply QCD directly
to the study of these hadronic processes, it is definitely feasible to 
extract model--independent information on the form factors of QCD currents
by exploiting and implementing the known features of the underlying theory,
such as symmetries or dynamic behaviour, providing a compelling framework
to work with.

\vspace*{0.8cm} 
\noindent
{\large \bf Acknowledgements}\par
\vspace{0.2cm}
\noindent 
We wish to thank Will Johns for correspondence on the published data
from E687. We also thank A. Pich for his comments in reading the
manuscript. Correspondence and conversations with D. Becirevic
are gratefully acknowledged.
This work has been supported in part by TMR, EC Contract No. 
ERB FMRX-CT98-0169, by MCYT (Spain) under grant
FPA-2001-3031, by 
DGESIC (Spain) under grant PB97-1401-C02-01, and by ERDF funds from
the European Commission.

\end{document}